\documentclass[aps,prl,reprint,showkeys,superscriptaddress,nobibnotes, floatfix]{revtex4-1}

\usepackage[pdftex]{graphicx}
\usepackage{epstopdf}
\usepackage{amsmath}
\usepackage{amssymb}

\begin{document}

\title{Symmetries of Chimera States}

\author{Felix P. Kemeth}
\affiliation{Physik-Department, Nonequilibrium Chemical Physics, Technische Universit\"{a}t M\"{u}nchen,
  James-Franck-Str. 1, D-85748 Garching, Germany}
\affiliation{Institute for Advanced Study - Technische Universit\"{a}t M\"{u}nchen,
  Lichtenbergstr. 2a, D-85748 Garching, Germany}
\author{Sindre W. Haugland}
\affiliation{Physik-Department, Nonequilibrium Chemical Physics, Technische Universit\"{a}t M\"{u}nchen,
  James-Franck-Str. 1, D-85748 Garching, Germany}
\affiliation{Institute for Advanced Study - Technische Universit\"{a}t M\"{u}nchen,
  Lichtenbergstr. 2a, D-85748 Garching, Germany}
\author{Katharina Krischer}
\affiliation{Physik-Department, Nonequilibrium Chemical Physics, Technische Universit\"{a}t M\"{u}nchen,
  James-Franck-Str. 1, D-85748 Garching, Germany}
\email[ ]{krischer@tum.de}

\begin{abstract}
  Symmetry broken states arise naturally in oscillatory networks.
  In this Letter, we investigate chaotic attractors in an ensemble of four mean-coupled Stuart-Landau oscillators with two oscillators being synchronized.
  We report that these states with partially broken symmetry, so-called chimera states, have different set-wise symmetries in the incoherent oscillators,
  and in particular some are and some are not invariant under a permutation symmetry on average.
  This allows for a classification of different chimera states in small networks.
  We conclude our report with a discussion of related states in spatially extended systems, which seem to inherit the symmetry properties of their counterparts in small networks.
\end{abstract}

\maketitle

It has been known for many years that symmetric coupling between identical oscillating units may lead to stable attracting sets with reduced symmetry, such as cluster states~\cite{Okuda1993}.
These states consist of two or more groups in which the individual oscillators behave identically.
In recent years, however, new states have been observed in which synchronized and incoherently oscillating groups coexist.
Since their identification in 2002~\cite{Kuramoto2002}, these so-called chimera states~\cite{Abrams2004} have attracted considerable interest and have been observed in numerous oscillatory systems,
many of them have been reviewed in recent literature~\cite{Panaggio2015,Kemeth2016}.
While the early studies considered large networks or spatially extended systems,
they also appear in small systems of just four units, in which two oscillators are synchronized and two are desynchronized~\cite{Ashwin2015,Hart2016,Panaggio2016,Bick2017}.
In systems of phase oscillators, (weak) chimera states are characterized by different mean frequencies of the synchronized and desynchronized groups~\cite{Ashwin2015}.
The definition of weak chimeras related to symmetry breaking has been investigated in Ref.~\cite{Bick2017_2}.
For many theoretical studies~\cite{Sethia2014,Schmidt2015,Hart2016} and experiments~\cite{Schoenleber2014,Schmidt2014,Hart2016}, however, the dynamics are dominated by amplitude fluctuations,
rendering a phase reduction, and in turn a classification based on phase dynamics, impossible. 

In this Letter, we investigate different kinds of chimera states in small networks of coupled oscillators even beyond phase oscillator systems.
Different states are distinguished using the set-wise symmetries of the attracting manifold, which can be determined with symmetry detectives~\cite{Barany1993,Dellnitz1994}.
We apply this method to various dynamical states observed in mean-coupled Stuart-Landau oscillators, and relate those states to different chimera states reported in recent literature.
Finally, we show and discuss how our results extend to larger networks and spatially extended systems.

Given a dynamical system
\begin{equation}
  \label{eq:sys_eq}
  \dot{\vec{x}} = \vec{f}\left(\vec{x}\right),
\end{equation}
then this system is invariant under the operation $\gamma$ if
\begin{equation}
  \label{eq:equi_sys}
  \vec{f}\left(\gamma \vec{x}\right) = \gamma \vec{f}\left(\vec{x}\right).
\end{equation}
The group $\{\gamma\}$ fulfilling Eq.~\eqref{eq:equi_sys} is called the symmetry group $\Gamma$ and system~\eqref{eq:sys_eq} is said be be $\Gamma$-equivariant~\cite{Golubitsky2003}.
However, as mentioned above, solutions of Eq.~\eqref{eq:sys_eq} are not necessarily invariant under the same symmetry group $\Gamma$,
i.e.\ the symmetry of solutions can be broken.
Let $\vec{x}$ be a solution of system~\eqref{eq:sys_eq},
then the group of transformations that leave $\vec{x}$ invariant,
\begin{equation*}
  \Sigma_{\vec{x}} = \left\{ \gamma \in \Gamma : \gamma \vec{x} = \vec{x}\right\},
\end{equation*}
is called the isotropy subgroup of $\vec{x}$.
Note that $\Sigma_{\vec{x}} \subseteq \Gamma$.

Even turbulent or spatio-temporally chaotic states may exhibit some symmetries in their time-averaged dynamics~\cite{Chossat1988,Gluckman1993,Campbell1987}.
Such symmetries are related to the set-wise symmetry of the attractor, that is, the group of symmetry operations that leave the whole attractor invariant.
If the dimension of the phase space is three or less, such symmetries can be observed visually, see for example Ref.~\cite{Field2009}.
For higher-dimensional systems, Barany et al.\ proposed so-called symmetry detectives~\cite{Barany1993}.
The idea is to transform the task of finding the symmetry group of a set $A$ in space $V$ to
finding the symmetries of a single point $K_A$ in some auxiliary space $\tilde{V}$~\cite{Barany1993, Golubitsky2003}.
This can be achieved by projecting the set $A$ through a $\Gamma$-equivariant map $\phi:V\rightarrow \tilde{V}$.
Then, $K_A$ can be expressed as
\begin{equation*}
  K_A = \underset{T \rightarrow \infty}{\lim} \frac{1}{T} \int_{0}^{T} \phi\left(x\left(t\right)\right)dt
\end{equation*}
for continuous dynamical systems~\cite{Dellnitz1994}.
$\phi: V \rightarrow \tilde{V}$ is called a detective with
\begin{equation*}
  \Sigma_{\phi\left(A\right)} = \Sigma_A,
\end{equation*}
if $\phi$ is $\Gamma$-equivariant and $\tilde{V}$ large enough, as explained in Ref.~\cite{Golubitsky2003}.
Once we mapped a trajectory of the dynamical system into the vector space $\tilde{V}$ using a detective function $\phi$ as described above,
we can estimate the symmetry $\Sigma_A$ of an attracting set by examining the isotropy subgroup $\Sigma(\omega)$ of $\omega = K_A \in \tilde{V}$~\cite{Ashwin1997}.
This can be achieved by taking $\omega_\gamma=K_{\gamma A}$ and computing the distances
\begin{equation*}
  t_\gamma = \lVert \omega_\gamma - \omega \rVert
\end{equation*}
for each symmetry operation $\gamma \in \Gamma$.
The isotropy group $\Sigma(\omega)$ is thus the set of all $\gamma$ for which $t_\gamma\approx 0$.
This is in contrast to the instantaneous symmetry of a solution, which is the intersection of the isotropy groups $\Sigma(x)$ at every position of the attractor,
\begin{equation*}
  \Sigma_{\mathrm{instant}} = \underset{x\in A}{\cap} \Sigma(x).
\end{equation*}
Using these two estimates, one can calculate the instantaneous and set-wise symmetries of an attracting manifold.

We apply this method to different solutions of $N=4$ Stuart-Landau oscillators, linearly coupled through the ensemble average
\begin{equation}
  \partial_t W_k = W_k - \left(1+ic_2\right)\left|W_k\right|^2 W_k + \kappa \left(\frac{1}{N} \sum_{j=1}^N W_j - W_k\right),
  \label{eq:sl_ens}
\end{equation}
with the complex variables $W_k$, $k=1,\dots,N$, the shear parameter $c_2\in\mathbb{R}$ and the coupling constant $\kappa=\alpha+i\beta$, $\alpha,\beta\in\mathbb{R}$.
This is motivated by the fact that this system shows chimera-like dynamics for large $N$~\cite{Sethia2014}.
Here, we fix $c_2=2$ and $\beta=-0.7$ and keep $\alpha$ as a tunable parameter.
Since the $W_k$ are complex, Eq.~\eqref{eq:sl_ens} describes the temporal evolution in $\mathbb{C}^4\cong\mathbb{R}^8$.
Furthermore, note that Eq.~\eqref{eq:sl_ens} is invariant under a permutation of the indices, $\mathbf{S}_4$, and a phase shift $W\rightarrow W\text{exp}\left(i\theta\right)$.
The latter can be eliminated using the transformed variables $R_k=\left|W_k\right|$, $k=1,\dots,4$, and $\Delta\theta_{k+1,k} = \theta_{k+1}-\theta_k=\angle W_{k+1}-\angle W_{k}$, $k=1,\dots,3$,
describing the dynamics in a seven-dimensional phase space ($\mathbb{R}_+^4 \times \mathbf{T}^3$, with $\mathbf{T} = \mathbb{R}/2\pi \mathbb{Z}$).
Thus, a limit cycle in the original variables, Eq.~\eqref{eq:sl_ens}, corresponds to a fixed point in the new amplitude and phase-difference variables.
See Supplement for the equations in these new variables and for details on the numerical methods used to integrate them~\cite{supplement}.

As shown in Ref.~\cite{Tchistiakov1996} for systems with the symbol permutation symmetry $\mathbf{S}_N$, one can use the ring group $\mathbf{R}_\Gamma$ as auxiliary space $\tilde{V}$ with the polynomial detective
\begin{equation*}
  \phi_k(\vec{x}) = p\left(\gamma_k^{-1} \vec{x}\right) \; , \; p = x_1 x_2^2 \dots x_{N-1}^{N-1},
\end{equation*}
with $k=\nobreak1,\dots,\left|\mathbf{S}_N\right|$ and $\gamma_k^{-1}$ being the inverse of $\gamma_k\in\nobreak\mathbf{S}_N$.
That is, for four globally coupled oscillators with $\mathbf{S}_4$ symmetry of order $\left|\mathbf{S}_4 \right|=24$, a possible choice for a symmetry detective is
\begin{equation*}
  \phi_k(\vec{x}) = p\left(\gamma_k^{-1} \vec{x}\right) \; , \; p = x_1 x_2^2 x_{3}^{3}  \; \rightarrow \; \vec{\phi}\left(\vec{x}\right) = 
  \begin{pmatrix} x_1 x_2^2 x_3^3 \\ x_2 x_1^2 x_3^3 \\ \vdots \\ x_4 x_3^2 x_2^3
  \end{pmatrix},
\end{equation*}
which we adopt in this letter, although other choices of $\phi$ are also possible~\cite{Ashwin1997}.
Here, we take the real parts of our complex time series $W_k$ as input $x_k$.

\begin{figure}[htb]
  \centering
  \includegraphics{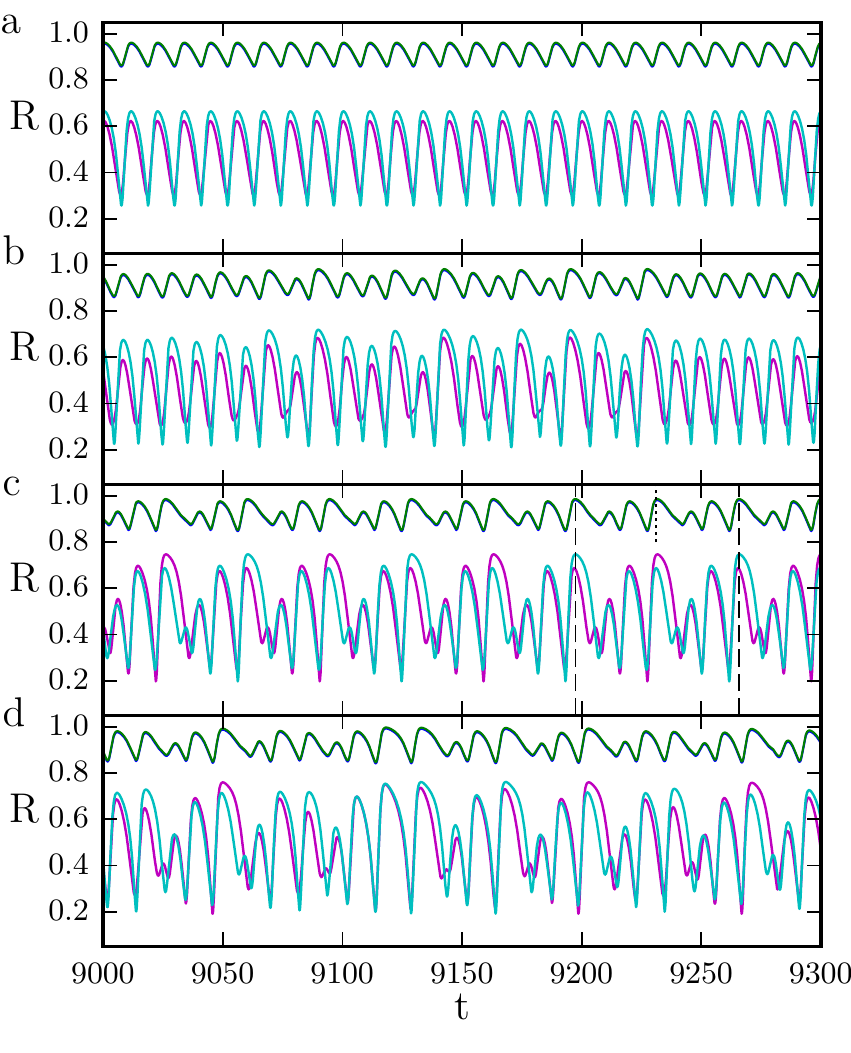}
  \caption{Time series of the amplitudes of the four oscillators of (a) a periodic orbit for $\alpha=0.85$,
    (b) an asymmetric chimera state at $\alpha=0.83764$,
    (c) a periodic orbit with phase shift symmetry at $\alpha=0.8376$ and
    (d) a symmetric chimera state at $\alpha=0.8365$.
    The other parameters are $\beta=-0.7$ and $c_2=2.0$.
    Note that two oscillators (here green and blue) are always synchronized for these parameter values and thus form only one curve.
    The vertical lines in (c) indicate the period of the desynchronized oscillators (dashed) and synchronized oscillators (dotted), respectively.}
\label{fig:4time_series}
\end{figure}

We start our considerations from a stable fixed-point solution with $R_1=R_2>R_3=R_4$, $\Delta\theta_{21}=0=\Delta\theta_{43}$ and $\Delta\theta_{32}\neq 0$ (first state in Table~\ref{table:1}),
with broken symmetry $\mathbf{S}_2\times\mathbf{S}_2$,
which can be obtained analytically (See Supplement for its analytic derivation~\cite{supplement}).
Reducing $\alpha$ leads to a supercritical Hopf bifurcation, where the fixed point solution becomes unstable and a stable periodic orbit is created.
That is, the amplitudes and the phase difference $\Delta\theta_{32}$ start to oscillate.
Further changing $\alpha$ leads to a pitchfork bifurcation, resulting in a reduced symmetry of the periodic orbit in which only two oscillators remain synchronized, $R_1=R_2$ and $\Delta\theta_{21}=0$,
the other two oscillators now having different amplitudes $R_4<R_3<R_1$ and phases, $\Delta \theta_{43}\neq 0$.
\begin{table}
\begin{tabular}{ l  c  r  c}
  \hline
  \hline
  State & $\alpha$-range & Symmetry & Index\\
  \hline
  2-2 FP & $> 0.8760$ & $\mathbf{S}_2^i\times\mathbf{S}_2^i $ &\\
  2-2 PO & $0.8760$ to $0.8562$ & $\mathbf{S}_2^i\times\mathbf{S}_2^i $  &\\
  2-1-1 PO & $0.8562$ to $0.8381$ & $ \mathbf{S}_2^i $  & a\\
  2-1-1 Chaos & $0.8381$ to $0.8376$ & $\mathbf{S}_2^i$ & b\\
  2-1-1 P6O & $0.8376$ to $0.8374$ & $\mathbf{S}_2^i\times\mathbf{\Xi}_2$  & c\\
  2-1-1 Chaos & $<0.8372$  & $\mathbf{S}_2^i\times\mathbf{S}_2^a$  & d\\
  \hline
  \hline
\end{tabular}
\caption{Different states observed in the system of four mean-coupled Stuart-Landau oscillators for $\beta=-0.7$ and $c_2=2$.
  FP denotes fixed point solution in the amplitude and phase difference variables, PO periodic orbits, Chaos indicates chaotic dynamics (chimeras) and P6O a period-6 orbit.
  The numbers indicate the number of synchronized oscillators, and the indices a-d correspond to the regions in Fig.~\ref{fig:feigenbaum} and the time series in Fig.~\ref{fig:4time_series}.}
\label{table:1}
\end{table}
The time series of the amplitudes of such a state are depicted in Fig.~\ref{fig:4time_series}(a).
This limit cycle gets destroyed through a period-doubling bifurcation at which a stable period-2 orbit is created.

\begin{figure}[htb]
  \centering
  \includegraphics{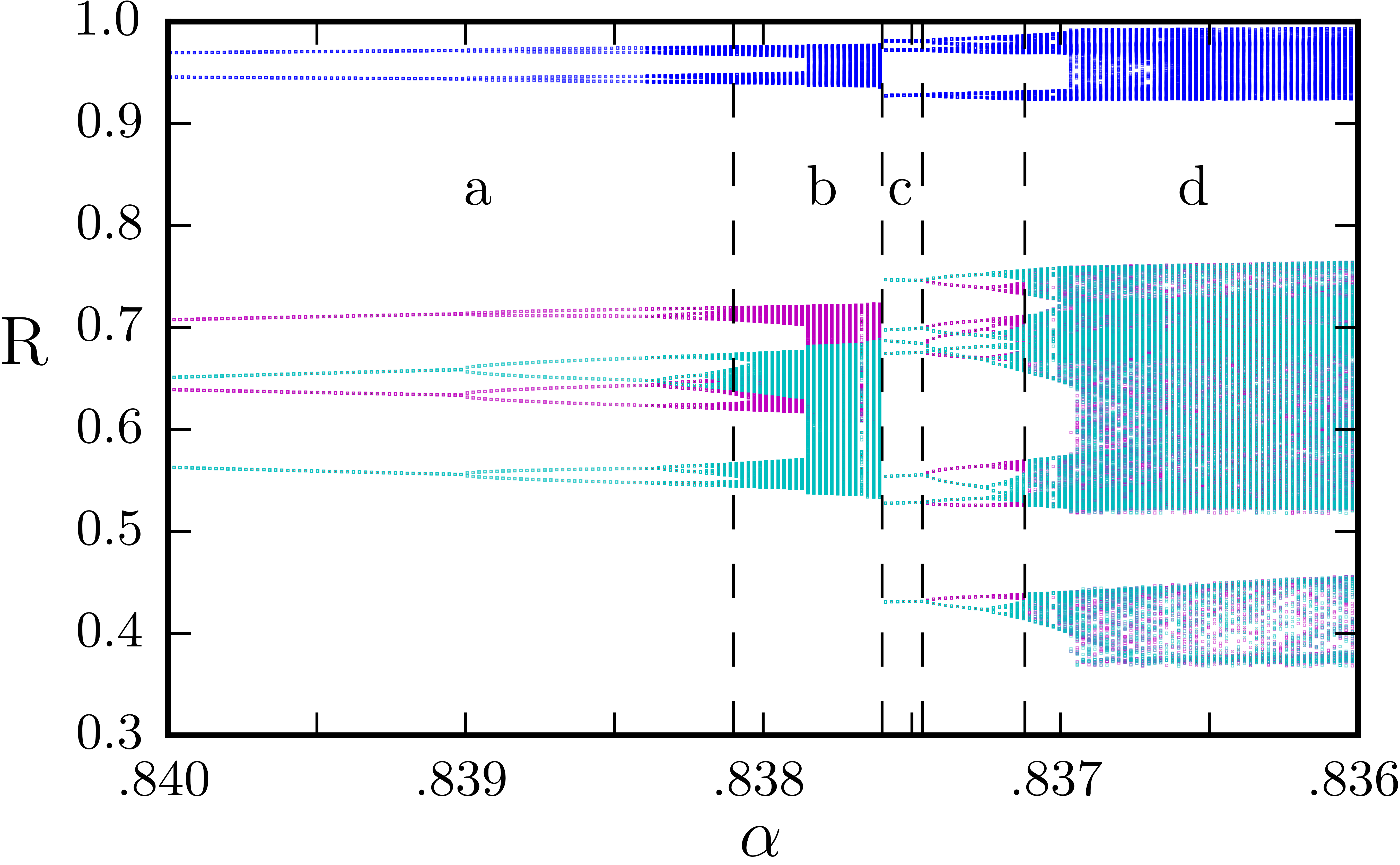}
  \caption{Poincar\'{e} map recording the maxima of the amplitudes of the individual oscillators for $0.84\geq\alpha\geq 0.836$ and $\beta=-0.7$.
    Region (a) marks the parameter range in which periodic orbits are observed, starting with a a period-doubled state,
    (b) indicates the existence of asymmetric chimera states,
    (c) denotes the region in which periodic orbits with discrete rotating wave symmetry exist,
    and for $\alpha$ values in region (d) symmetric chimera states are observed.
    The corresponding time series of the amplitudes are shown in Fig.~\ref{fig:4time_series}.}
\label{fig:feigenbaum}
\end{figure}

\begin{figure}[htb]
  \centering
  \includegraphics{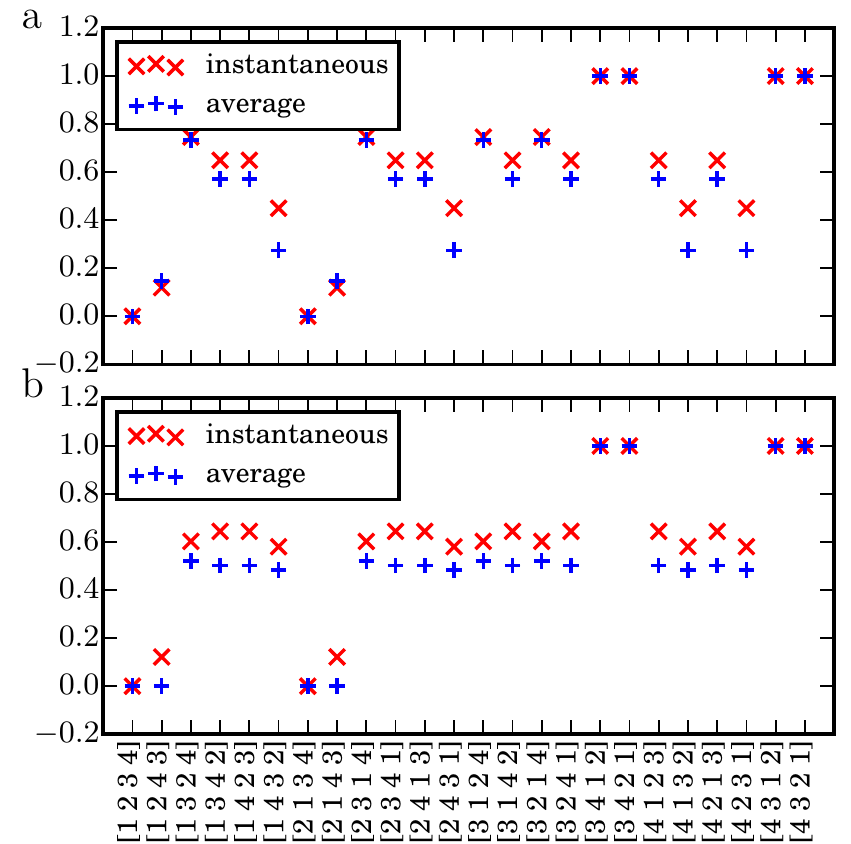}
  \caption{The distances $t_\gamma$ for the symmetry operations $\gamma\in \mathbf{S}_4$.
    $t_\gamma\approx 0$ indicates the instantaneous and average symmetries of (a) the asymmetric chimera states at $\alpha=0.83764$ and (b) the symmetric chimera state at $\alpha=0.8365$,
  suggesting that the asymmetric chimera is invariant under the actions of $\mathbf{S}_2^i$, and the symmetric chimera under the actions of $\mathbf{S}_2^i \times \mathbf{S}_2^a$.}
\label{fig:sym_detect}
\end{figure}

Periodic solutions are best visualized using a Poincar\'{e} section, recording the dynamical states at discrete points in time~\cite{Strogatz2015}.
Here, we use a representation, as shown in Fig.~\ref{fig:feigenbaum}, plotting only the value of each amplitude $R_k$ when it becomes maximal.
Note that a simple periodic orbit appears as a single point per oscillator, whereas two points per oscillator in the Poincar\'{e} section indicate a period-2 orbit, see left side of Fig.~\ref{fig:feigenbaum}.
This reduces the dimension of the trajectories and simplifies the analysis.
In particular, as shown in Fig.~\ref{fig:feigenbaum}, the period-2 orbit bifurcates into a period-4 orbit when $\alpha$ is reduced.
This subsequently bifurcates into a period-8 orbit and so forth.
In other words, one observes a cascade of infinitely many period-doubling bifurcations~\cite{Feigenbaum1978}, leading to a chaotic state (region b in Fig.~\ref{fig:feigenbaum}).
The time series of the amplitudes of such a chaotic attractor are depicted in Fig.~\ref{fig:4time_series}(b).
It is important to notice that the phase and the amplitude difference of two of the oscillators is zero, indicating that, although the total dynamics are chaotic, they are synchronized.
Furthermore, this chimera state is not invariant under a permutation of the third and fourth oscillator.
In other words, the two incoherent oscillators are not symmetric.
This can be verified using symmetry detectives, as shown in Fig.~\ref{fig:sym_detect}(a).
There, one can see that the distances $t_\gamma$ are non-zero when $\gamma$ involves a permutation of the two incoherent oscillators.
In other words, the underlying chaotic attractor has an $\mathbf{S}_2^i$ symmetry  in the two synchronized oscillators only, with the superscript $i$ indicating that the symmetry is instantaneous.

Further reducing $\alpha$ destroys the chimera state, yielding again a periodic state (region c in Fig.~\ref{fig:feigenbaum}), with the time series shown in Fig.~\ref{fig:4time_series}(c).
From the amplitude time series one can observe that the two desynchronized oscillators perform the same oscillations but with a constant phase shift.
Such symmetry is called a phase-shift symmetry or discrete rotating wave~\cite{Golubitsky2003,Golubitsky2015}, reminiscent of the rotating waves observed in Ref.~\cite{Schmidt2015}.
Denoting the phase-shift symmetry of the two nonsynchronized oscillators with $\mathbf{\Xi}_2$, this state has an isotropy subgroup $\mathbf{S}_2^i \times \mathbf{\Xi}_2$.
Furthermore, it is worth mentioning that, due to the rotating-wave symmetry, the frequency of the oscillation in the amplitudes of the synchronized oscillators is twice the frequency of the
desynchronized oscillators. This is reminiscent of the weak chimera states reported in Ref.~\cite{Ashwin2015}, which are periodic but have different mean frequencies in the individual oscillators.

Further decreasing $\alpha$ first leads to a pitchfork bifurcation in which orbits with reduced symmetries are born,
similar to the symmetry-decreasing bifurcations reported in Ref.~\cite{vanderWeele1992}.
After another cascade of period-doubling bifurcations, one again obtains chaotic dynamics, see the time series in Fig.~\ref{fig:4time_series}(d).
Surprisingly, and opposed to the chimera state described above, this attractor is symmetric under a permutation of the two desynchronized oscillators.
That the attracting manifold is indeed invariant under such a symmetry operation can be verified using the symmetry detectives mentioned above,
with the distances $t_\gamma$ shown in Fig.~\ref{fig:sym_detect}(b).
Note that a distance close to zero indicates an invariance under the respective group action, whereas $t_\gamma \neq 0$ indicates the absence of such a symmetry.
Thus the symmetric chimera state has an $\mathbf{S}_2^i\times\mathbf{S}_2^a$ symmetry, different from asymmetric chimera states with sole $\mathbf{S}_2^i$ symmetry.
For a summary of the states discussed so far, see table~\ref{table:1}.

Calculating the symmetry detectives of the four coupled opto-electronic oscillators reported in Ref.~\cite{Hart2016},
we find that also those states have an $\mathbf{S}_2^i\times\mathbf{S}_2^a$ symmetry,
and can thus be identified as symmetric chimera states.


In order to see if the states discussed above persist for larger ensembles of oscillators and under the influence of diffusion, we modify Eq.~\eqref{eq:sl_ens} by adding a
diffusive coupling to the ensemble,
\begin{align*}
  \partial_t W(x,t) &= W(x,t) - \left(1+ic_2\right)\left|W(x,t)\right|^2 W(x,t) \nonumber\\
  &+ \kappa \left(\frac{1}{L} \int_L W(x,t)dx - W(x,t)\right) + \partial_{xx} W(x,t)
\end{align*}
yielding a version of the complex Ginzburg-Landau equation with one spatial dimension $\mathrm{x}$ and linear global coupling, indicated through the spatial integral~\cite{GarciaMorales2012,Battogtokh1996}.
\begin{figure}
  \centering
  \includegraphics{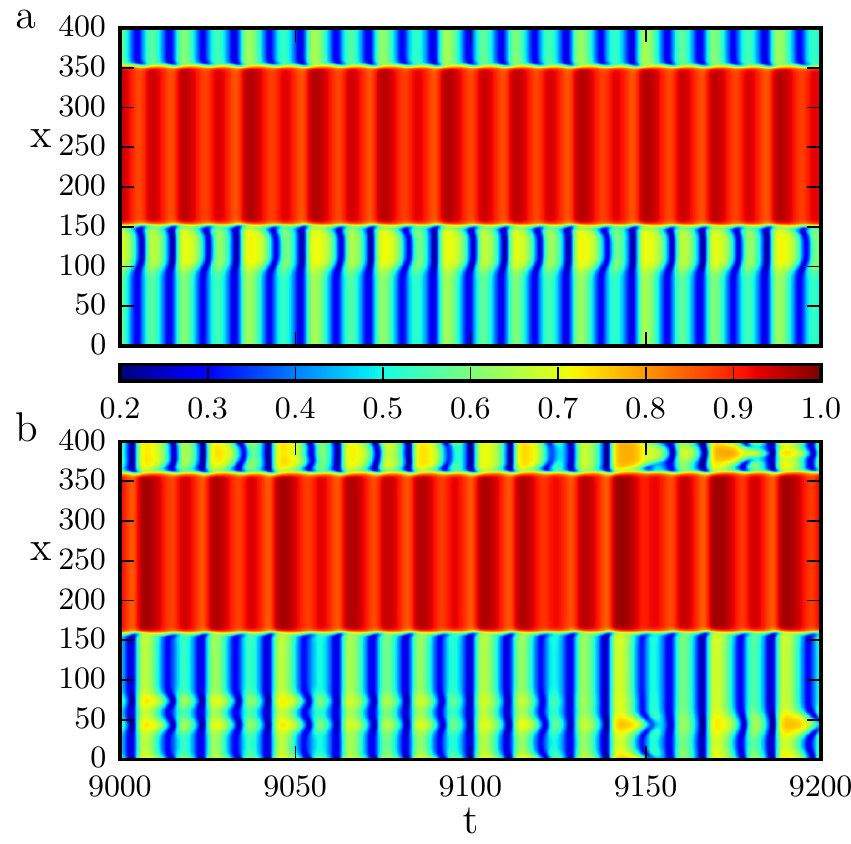}
  \caption{Space-time plot of the (a) asymmetric chimera in the spatially extended system with $L=400$, $\alpha=0.8304$, $\beta=-0.7$ and $c_2=2$.
    The asymmetry arises through the two clusters with small but different amplitudes (blue-ish and yellow-ish color in plot (a)).
    (b) Symmetric chimera in the spatially extended system with $L=400$, $\alpha=0.828$, $\beta=-0.7$ and $c_2=2$.
    The color encodes the absolute value of $W$.}
\label{fig:extended_chimeras}
\end{figure}
Numerically solving this system on a domain of length $L=400$ and periodic boundary conditions,
one obtains chaotic states resembling the asymmetric chimera (see Fig.~\ref{fig:extended_chimeras}(a)),
and the symmetric chimera (see Fig.~\ref{fig:extended_chimeras}(b)) for different parameter values.
%
%
Interestingly, in the spatially extended system the asymmetric chimera state of the four-oscillator network conserves its low-dimensional dynamics, manifesting itself in a three-cluster state with temporally chaotic behavior. A comparison of time-series recorded at a position within each of the three clusters and those shown in Fig.~\ref{fig:4time_series}(b) is given in the Supplement and substantiates the correspondence of these states~\cite{supplement}. In contrast, the symmetric chimera state transforms into a spatio-temporal chimera state with a synchronized, temporally chaotic cluster and a spatially incoherent, temporally chaotic region, as can be seen in Fig.~\ref{fig:extended_chimeras}(b).
Corresponding time series of this apparently \emph{extensive chimera state} are again displayed together with its low-dimensional counterparts in the Supplement~\cite{supplement}.
Note that the $\alpha$ values at which those states arise are slightly shifted compared to the corresponding states observed in the four-oscillator system.
This is an effect of the diffusion and the different sizes of the clusters.

To summarize our results, we find different kinds of symmetry-broken states in a system of four globally coupled oscillators.
In particular, we report chaotic states with $\mathbf{S}_2^i$ symmetry, which we dub asymmetric chimera states, states with $\mathbf{S}_2^i\times\mathbf{S}_2^a$ symmetry,
which we call symmetric chimera states,
and deterministic periodic orbits with $\mathbf{S}_2^i\times\mathbf{\Xi}_2$ symmetry.
The latter resemble weak chimeras as defined for phase oscillators, whereas we could show that the symmetric chimera states also exist in a system reported in Ref.~\cite{Hart2016}.
The discrimination based on the symmetries of the incoherent oscillators may, as we hope, facilitate our understanding of intricate dynamics such as chimera states, and may help to further classify them.
In addition, such minimal chimera states in small networks may further reveal insights into dynamics of larger, and even spatially extended, systems of oscillators, which, as we have seen,
maintain certain properties of their minimal relatives.
In addition, our studies revealed apparent weaknesses in the concept of chimeras in small systems,
since some of their spatially-extended counterparts remain spatially synchronized exhibiting low-dimensional dynamics,
while some other develop extensive spatio-temporal incoherence.
This directly relates to the question of how the dynamics changes from small systems to large ensembles,
which is, in our opinion, an important and challenging question for future research.

\acknowledgments{The authors thank Vladimir Garc\'{i}a-Morales, Matthias Wolfrum, Oliver Junge, Christian Bick and Munir Salman for fruitful discussions.
  Financial support from the Deutsche Forschungsgemeinschaft (Grant no. KR 1189/18-1), the Institute of Advanced Study - Technische Universit\"{a}t M\"{u}nchen, funded by
  the German Excellence Initiative,
  and the Studienstiftung des deutschen Volkes is gratefully acknowledged.}
\bibliography{lit}

\begin{thebibliography}{32}%
\makeatletter
\providecommand \@ifxundefined [1]{%
 \@ifx{#1\undefined}
}%
\providecommand \@ifnum [1]{%
 \ifnum #1\expandafter \@firstoftwo
 \else \expandafter \@secondoftwo
 \fi
}%
\providecommand \@ifx [1]{%
 \ifx #1\expandafter \@firstoftwo
 \else \expandafter \@secondoftwo
 \fi
}%
\providecommand \natexlab [1]{#1}%
\providecommand \enquote  [1]{``#1''}%
\providecommand \bibnamefont  [1]{#1}%
\providecommand \bibfnamefont [1]{#1}%
\providecommand \citenamefont [1]{#1}%
\providecommand \href@noop [0]{\@secondoftwo}%
\providecommand \href [0]{\begingroup \@sanitize@url \@href}%
\providecommand \@href[1]{\@@startlink{#1}\@@href}%
\providecommand \@@href[1]{\endgroup#1\@@endlink}%
\providecommand \@sanitize@url [0]{\catcode `\\12\catcode `\$12\catcode
  `\&12\catcode `\#12\catcode `\^12\catcode `\_12\catcode `\%12\relax}%
\providecommand \@@startlink[1]{}%
\providecommand \@@endlink[0]{}%
\providecommand \url  [0]{\begingroup\@sanitize@url \@url }%
\providecommand \@url [1]{\endgroup\@href {#1}{\urlprefix }}%
\providecommand \urlprefix  [0]{URL }%
\providecommand \Eprint [0]{\href }%
\providecommand \doibase [0]{http://dx.doi.org/}%
\providecommand \selectlanguage [0]{\@gobble}%
\providecommand \bibinfo  [0]{\@secondoftwo}%
\providecommand \bibfield  [0]{\@secondoftwo}%
\providecommand \translation [1]{[#1]}%
\providecommand \BibitemOpen [0]{}%
\providecommand \bibitemStop [0]{}%
\providecommand \bibitemNoStop [0]{.\EOS\space}%
\providecommand \EOS [0]{\spacefactor3000\relax}%
\providecommand \BibitemShut  [1]{\csname bibitem#1\endcsname}%
\let\auto@bib@innerbib\@empty
\bibitem [{\citenamefont {Okuda}(1993)}]{Okuda1993}%
  \BibitemOpen
  \bibfield  {author} {\bibinfo {author} {\bibfnamefont {K.}~\bibnamefont
  {Okuda}},\ }\href {\doibase 10.1016/0167-2789(93)90121-g} {\bibfield
  {journal} {\bibinfo  {journal} {Physica D: Nonlinear Phenomena}\ }\textbf
  {\bibinfo {volume} {63}},\ \bibinfo {pages} {424} (\bibinfo {year}
  {1993})}\BibitemShut {NoStop}%
\bibitem [{\citenamefont {Kuramoto}\ and\ \citenamefont
  {Battogtokh}(2002)}]{Kuramoto2002}%
  \BibitemOpen
  \bibfield  {author} {\bibinfo {author} {\bibfnamefont {Y.}~\bibnamefont
  {Kuramoto}}\ and\ \bibinfo {author} {\bibfnamefont {D.}~\bibnamefont
  {Battogtokh}},\ }\href@noop {} {\bibfield  {journal} {\bibinfo  {journal}
  {Nonlinear Phenom. Complex Syst.}\ }\textbf {\bibinfo {volume} {5}},\
  \bibinfo {pages} {380} (\bibinfo {year} {2002})}\BibitemShut {NoStop}%
\bibitem [{\citenamefont {Abrams}\ and\ \citenamefont
  {Strogatz}(2004)}]{Abrams2004}%
  \BibitemOpen
  \bibfield  {author} {\bibinfo {author} {\bibfnamefont {D.~M.}\ \bibnamefont
  {Abrams}}\ and\ \bibinfo {author} {\bibfnamefont {S.~H.}\ \bibnamefont
  {Strogatz}},\ }\href {\doibase 10.1103/PhysRevLett.93.174102} {\bibfield
  {journal} {\bibinfo  {journal} {Phys. Rev. Lett.}\ }\textbf {\bibinfo
  {volume} {93}},\ \bibinfo {pages} {174102} (\bibinfo {year}
  {2004})}\BibitemShut {NoStop}%
\bibitem [{\citenamefont {Panaggio}\ and\ \citenamefont
  {Abrams}(2015)}]{Panaggio2015}%
  \BibitemOpen
  \bibfield  {author} {\bibinfo {author} {\bibfnamefont {M.~J.}\ \bibnamefont
  {Panaggio}}\ and\ \bibinfo {author} {\bibfnamefont {D.~M.}\ \bibnamefont
  {Abrams}},\ }\href {http://stacks.iop.org/0951-7715/28/i=3/a=R67} {\bibfield
  {journal} {\bibinfo  {journal} {Nonlinearity}\ }\textbf {\bibinfo {volume}
  {28}},\ \bibinfo {pages} {R67} (\bibinfo {year} {2015})}\BibitemShut
  {NoStop}%
\bibitem [{\citenamefont {Kemeth}\ \emph {et~al.}(2016)\citenamefont {Kemeth},
  \citenamefont {Haugland}, \citenamefont {Schmidt}, \citenamefont
  {Kevrekidis},\ and\ \citenamefont {Krischer}}]{Kemeth2016}%
  \BibitemOpen
  \bibfield  {author} {\bibinfo {author} {\bibfnamefont {F.~P.}\ \bibnamefont
  {Kemeth}}, \bibinfo {author} {\bibfnamefont {S.~W.}\ \bibnamefont
  {Haugland}}, \bibinfo {author} {\bibfnamefont {L.}~\bibnamefont {Schmidt}},
  \bibinfo {author} {\bibfnamefont {I.~G.}\ \bibnamefont {Kevrekidis}}, \ and\
  \bibinfo {author} {\bibfnamefont {K.}~\bibnamefont {Krischer}},\ }\href
  {\doibase 10.1063/1.4959804} {\bibfield  {journal} {\bibinfo  {journal}
  {Chaos: An Interdisciplinary Journal of Nonlinear Science}\ }\textbf
  {\bibinfo {volume} {26}},\ \bibinfo {pages} {094815} (\bibinfo {year}
  {2016})},\ \Eprint {http://arxiv.org/abs/https://doi.org/10.1063/1.4959804}
  {https://doi.org/10.1063/1.4959804} \BibitemShut {NoStop}%
\bibitem [{\citenamefont {Ashwin}\ and\ \citenamefont
  {Burylko}(2015)}]{Ashwin2015}%
  \BibitemOpen
  \bibfield  {author} {\bibinfo {author} {\bibfnamefont {P.}~\bibnamefont
  {Ashwin}}\ and\ \bibinfo {author} {\bibfnamefont {O.}~\bibnamefont
  {Burylko}},\ }\href {\doibase 10.1063/1.4905197} {\bibfield  {journal}
  {\bibinfo  {journal} {Chaos: An Interdisciplinary Journal of Nonlinear
  Science}\ }\textbf {\bibinfo {volume} {25}},\ \bibinfo {pages} {013106}
  (\bibinfo {year} {2015})},\ \Eprint
  {http://arxiv.org/abs/https://doi.org/10.1063/1.4905197}
  {https://doi.org/10.1063/1.4905197} \BibitemShut {NoStop}%
\bibitem [{\citenamefont {Hart}\ \emph {et~al.}(2016)\citenamefont {Hart},
  \citenamefont {Bansal}, \citenamefont {Murphy},\ and\ \citenamefont
  {Roy}}]{Hart2016}%
  \BibitemOpen
  \bibfield  {author} {\bibinfo {author} {\bibfnamefont {J.~D.}\ \bibnamefont
  {Hart}}, \bibinfo {author} {\bibfnamefont {K.}~\bibnamefont {Bansal}},
  \bibinfo {author} {\bibfnamefont {T.~E.}\ \bibnamefont {Murphy}}, \ and\
  \bibinfo {author} {\bibfnamefont {R.}~\bibnamefont {Roy}},\ }\href {\doibase
  10.1063/1.4953662} {\bibfield  {journal} {\bibinfo  {journal} {Chaos: An
  Interdisciplinary Journal of Nonlinear Science}\ }\textbf {\bibinfo {volume}
  {26}},\ \bibinfo {pages} {094801} (\bibinfo {year} {2016})},\ \Eprint
  {http://arxiv.org/abs/https://doi.org/10.1063/1.4953662}
  {https://doi.org/10.1063/1.4953662} \BibitemShut {NoStop}%
\bibitem [{\citenamefont {Panaggio}\ \emph {et~al.}(2016)\citenamefont
  {Panaggio}, \citenamefont {Abrams}, \citenamefont {Ashwin},\ and\
  \citenamefont {Laing}}]{Panaggio2016}%
  \BibitemOpen
  \bibfield  {author} {\bibinfo {author} {\bibfnamefont {M.~J.}\ \bibnamefont
  {Panaggio}}, \bibinfo {author} {\bibfnamefont {D.~M.}\ \bibnamefont
  {Abrams}}, \bibinfo {author} {\bibfnamefont {P.}~\bibnamefont {Ashwin}}, \
  and\ \bibinfo {author} {\bibfnamefont {C.~R.}\ \bibnamefont {Laing}},\ }\href
  {\doibase 10.1103/PhysRevE.93.012218} {\bibfield  {journal} {\bibinfo
  {journal} {Phys. Rev. E}\ }\textbf {\bibinfo {volume} {93}},\ \bibinfo
  {pages} {012218} (\bibinfo {year} {2016})}\BibitemShut {NoStop}%
\bibitem [{\citenamefont {Bick}\ \emph {et~al.}(2017)\citenamefont {Bick},
  \citenamefont {Sebek},\ and\ \citenamefont {Kiss}}]{Bick2017}%
  \BibitemOpen
  \bibfield  {author} {\bibinfo {author} {\bibfnamefont {C.}~\bibnamefont
  {Bick}}, \bibinfo {author} {\bibfnamefont {M.}~\bibnamefont {Sebek}}, \ and\
  \bibinfo {author} {\bibfnamefont {I.~Z.}\ \bibnamefont {Kiss}},\ }\href
  {\doibase 10.1103/PhysRevLett.119.168301} {\bibfield  {journal} {\bibinfo
  {journal} {Phys. Rev. Lett.}\ }\textbf {\bibinfo {volume} {119}},\ \bibinfo
  {pages} {168301} (\bibinfo {year} {2017})}\BibitemShut {NoStop}%
\bibitem [{\citenamefont {Bick}(2017)}]{Bick2017_2}%
  \BibitemOpen
  \bibfield  {author} {\bibinfo {author} {\bibfnamefont {C.}~\bibnamefont
  {Bick}},\ }\href {\doibase 10.1007/s00332-016-9345-2} {\bibfield  {journal}
  {\bibinfo  {journal} {Journal of Nonlinear Science}\ }\textbf {\bibinfo
  {volume} {27}},\ \bibinfo {pages} {605} (\bibinfo {year} {2017})}\BibitemShut
  {NoStop}%
\bibitem [{\citenamefont {Sethia}\ and\ \citenamefont
  {Sen}(2014)}]{Sethia2014}%
  \BibitemOpen
  \bibfield  {author} {\bibinfo {author} {\bibfnamefont {G.~C.}\ \bibnamefont
  {Sethia}}\ and\ \bibinfo {author} {\bibfnamefont {A.}~\bibnamefont {Sen}},\
  }\href {\doibase 10.1103/PhysRevLett.112.144101} {\bibfield  {journal}
  {\bibinfo  {journal} {Phys. Rev. Lett.}\ }\textbf {\bibinfo {volume} {112}},\
  \bibinfo {pages} {144101} (\bibinfo {year} {2014})}\BibitemShut {NoStop}%
\bibitem [{\citenamefont {Schmidt}\ and\ \citenamefont
  {Krischer}(2015)}]{Schmidt2015}%
  \BibitemOpen
  \bibfield  {author} {\bibinfo {author} {\bibfnamefont {L.}~\bibnamefont
  {Schmidt}}\ and\ \bibinfo {author} {\bibfnamefont {K.}~\bibnamefont
  {Krischer}},\ }\href {\doibase 10.1063/1.4921727} {\bibfield  {journal}
  {\bibinfo  {journal} {Chaos: An Interdisciplinary Journal of Nonlinear
  Science}\ }\textbf {\bibinfo {volume} {25}},\ \bibinfo {pages} {064401}
  (\bibinfo {year} {2015})},\ \Eprint
  {http://arxiv.org/abs/https://doi.org/10.1063/1.4921727}
  {https://doi.org/10.1063/1.4921727} \BibitemShut {NoStop}%
\bibitem [{\citenamefont {Sch\"{o}nleber}\ \emph {et~al.}(2014)\citenamefont
  {Sch\"{o}nleber}, \citenamefont {Zensen}, \citenamefont {Heinrich},\ and\
  \citenamefont {Krischer}}]{Schoenleber2014}%
  \BibitemOpen
  \bibfield  {author} {\bibinfo {author} {\bibfnamefont {K.}~\bibnamefont
  {Sch\"{o}nleber}}, \bibinfo {author} {\bibfnamefont {C.}~\bibnamefont
  {Zensen}}, \bibinfo {author} {\bibfnamefont {A.}~\bibnamefont {Heinrich}}, \
  and\ \bibinfo {author} {\bibfnamefont {K.}~\bibnamefont {Krischer}},\ }\href
  {\doibase https://doi.org/10.1088/1367-2630/16/6/063024} {\bibfield
  {journal} {\bibinfo  {journal} {New Journal of Physics}\ }\textbf {\bibinfo
  {volume} {16}},\ \bibinfo {pages} {063024} (\bibinfo {year}
  {2014})}\BibitemShut {NoStop}%
\bibitem [{\citenamefont {Schmidt}\ \emph {et~al.}(2014)\citenamefont
  {Schmidt}, \citenamefont {Sch\"{o}nleber}, \citenamefont {Krischer},\ and\
  \citenamefont {Garc\'{i}a-Morales}}]{Schmidt2014}%
  \BibitemOpen
  \bibfield  {author} {\bibinfo {author} {\bibfnamefont {L.}~\bibnamefont
  {Schmidt}}, \bibinfo {author} {\bibfnamefont {K.}~\bibnamefont
  {Sch\"{o}nleber}}, \bibinfo {author} {\bibfnamefont {K.}~\bibnamefont
  {Krischer}}, \ and\ \bibinfo {author} {\bibfnamefont {V.}~\bibnamefont
  {Garc\'{i}a-Morales}},\ }\href {\doibase 10.1063/1.4858996} {\bibfield
  {journal} {\bibinfo  {journal} {Chaos: An Interdisciplinary Journal of
  Nonlinear Science}\ }\textbf {\bibinfo {volume} {24}},\ \bibinfo {pages}
  {013102} (\bibinfo {year} {2014})},\ \Eprint
  {http://arxiv.org/abs/https://doi.org/10.1063/1.4858996}
  {https://doi.org/10.1063/1.4858996} \BibitemShut {NoStop}%
\bibitem [{\citenamefont {Barany}\ \emph {et~al.}(1993)\citenamefont {Barany},
  \citenamefont {Dellnitz},\ and\ \citenamefont {Golubitsky}}]{Barany1993}%
  \BibitemOpen
  \bibfield  {author} {\bibinfo {author} {\bibfnamefont {E.}~\bibnamefont
  {Barany}}, \bibinfo {author} {\bibfnamefont {M.}~\bibnamefont {Dellnitz}}, \
  and\ \bibinfo {author} {\bibfnamefont {M.}~\bibnamefont {Golubitsky}},\
  }\href {\doibase https://doi.org/10.1016/0167-2789(93)90198-A} {\bibfield
  {journal} {\bibinfo  {journal} {Physica D: Nonlinear Phenomena}\ }\textbf
  {\bibinfo {volume} {67}},\ \bibinfo {pages} {66} (\bibinfo {year}
  {1993})}\BibitemShut {NoStop}%
\bibitem [{\citenamefont {Dellnitz}\ \emph {et~al.}(1994)\citenamefont
  {Dellnitz}, \citenamefont {Golubitsky},\ and\ \citenamefont
  {Nicol}}]{Dellnitz1994}%
  \BibitemOpen
  \bibfield  {author} {\bibinfo {author} {\bibfnamefont {M.}~\bibnamefont
  {Dellnitz}}, \bibinfo {author} {\bibfnamefont {M.}~\bibnamefont
  {Golubitsky}}, \ and\ \bibinfo {author} {\bibfnamefont {M.}~\bibnamefont
  {Nicol}},\ }\enquote {\bibinfo {title} {Symmetry of attractors and the
  karhunen-lo{\`e}ve decomposition},}\ in\ \href {\doibase
  10.1007/978-1-4612-0859-4_4} {\emph {\bibinfo {booktitle} {Trends and
  Perspectives in Applied Mathematics}}}\ (\bibinfo  {publisher} {Springer New
  York},\ \bibinfo {address} {New York, NY},\ \bibinfo {year} {1994})\ pp.\
  \bibinfo {pages} {73--108}\BibitemShut {NoStop}%
\bibitem [{\citenamefont {Golubitsky}\ and\ \citenamefont
  {Stuart}(2003)}]{Golubitsky2003}%
  \BibitemOpen
  \bibfield  {author} {\bibinfo {author} {\bibfnamefont {M.}~\bibnamefont
  {Golubitsky}}\ and\ \bibinfo {author} {\bibfnamefont {I.}~\bibnamefont
  {Stuart}},\ }\href@noop {} {\emph {\bibinfo {title} {The Symmetry
  Perspective: {F}rom Equilibrium to Chaos in Phase Space and Physical
  Space}}}\ (\bibinfo  {publisher} {Birk\"{a}user Verlag},\ \bibinfo {address}
  {Basel, Boston, Berlin},\ \bibinfo {year} {2003})\BibitemShut {NoStop}%
\bibitem [{\citenamefont {Chossat}\ and\ \citenamefont
  {Golubitsky}(1988)}]{Chossat1988}%
  \BibitemOpen
  \bibfield  {author} {\bibinfo {author} {\bibfnamefont {P.}~\bibnamefont
  {Chossat}}\ and\ \bibinfo {author} {\bibfnamefont {M.}~\bibnamefont
  {Golubitsky}},\ }\href {\doibase
  https://doi.org/10.1016/0167-2789(88)90066-8} {\bibfield  {journal} {\bibinfo
   {journal} {Physica D: Nonlinear Phenomena}\ }\textbf {\bibinfo {volume}
  {32}},\ \bibinfo {pages} {423} (\bibinfo {year} {1988})}\BibitemShut
  {NoStop}%
\bibitem [{\citenamefont {Gluckman}\ \emph {et~al.}(1993)\citenamefont
  {Gluckman}, \citenamefont {Marcq}, \citenamefont {Bridger},\ and\
  \citenamefont {Gollub}}]{Gluckman1993}%
  \BibitemOpen
  \bibfield  {author} {\bibinfo {author} {\bibfnamefont {B.~J.}\ \bibnamefont
  {Gluckman}}, \bibinfo {author} {\bibfnamefont {P.}~\bibnamefont {Marcq}},
  \bibinfo {author} {\bibfnamefont {J.}~\bibnamefont {Bridger}}, \ and\
  \bibinfo {author} {\bibfnamefont {J.~P.}\ \bibnamefont {Gollub}},\ }\href
  {\doibase 10.1103/PhysRevLett.71.2034} {\bibfield  {journal} {\bibinfo
  {journal} {Phys. Rev. Lett.}\ }\textbf {\bibinfo {volume} {71}},\ \bibinfo
  {pages} {2034} (\bibinfo {year} {1993})}\BibitemShut {NoStop}%
\bibitem [{\citenamefont {Campbell}(1987)}]{Campbell1987}%
  \BibitemOpen
  \bibfield  {author} {\bibinfo {author} {\bibfnamefont {D.~K.}\ \bibnamefont
  {Campbell}},\ }\href
  {http://permalink.lanl.gov/object/tr?what=info:lanl-repo/lareport/LA-UR-88-9072}
  {\bibfield  {journal} {\bibinfo  {journal} {Los Alamos science}\ }\textbf
  {\bibinfo {volume} {15}},\ \bibinfo {pages} {218} (\bibinfo {year}
  {1987})}\BibitemShut {NoStop}%
\bibitem [{\citenamefont {Field}\ and\ \citenamefont
  {Golubitsky}(2009)}]{Field2009}%
  \BibitemOpen
  \bibfield  {author} {\bibinfo {author} {\bibfnamefont {M.}~\bibnamefont
  {Field}}\ and\ \bibinfo {author} {\bibfnamefont {M.}~\bibnamefont
  {Golubitsky}},\ }\href {\doibase 10.1137/1.9780898717709} {\emph {\bibinfo
  {title} {Symmetry in Chaos}}},\ \bibinfo {edition} {2nd}\ ed.\ (\bibinfo
  {publisher} {Society for Industrial and Applied Mathematics},\ \bibinfo
  {year} {2009})\ \Eprint
  {http://arxiv.org/abs/http://epubs.siam.org/doi/pdf/10.1137/1.9780898717709}
  {http://epubs.siam.org/doi/pdf/10.1137/1.9780898717709} \BibitemShut
  {NoStop}%
\bibitem [{\citenamefont {Ashwin}\ and\ \citenamefont
  {Tomes}(1997)}]{Ashwin1997}%
  \BibitemOpen
  \bibfield  {author} {\bibinfo {author} {\bibfnamefont {P.}~\bibnamefont
  {Ashwin}}\ and\ \bibinfo {author} {\bibfnamefont {J.}~\bibnamefont {Tomes}},\
  }\href {\doibase https://doi.org/10.1016/S0167-2789(96)00176-5} {\bibfield
  {journal} {\bibinfo  {journal} {Physica D: Nonlinear Phenomena}\ }\textbf
  {\bibinfo {volume} {100}},\ \bibinfo {pages} {71} (\bibinfo {year}
  {1997})}\BibitemShut {NoStop}%
\bibitem [{sup()}]{supplement}%
  \BibitemOpen
  \href@noop {} {}\bibinfo {note} {See supplemental material at
  [supplement.pdf] for details on to the individual system and on the numerical
  methods used, including Refs. \cite{Cox2002,Hunter2007}}\BibitemShut
  {NoStop}%
\bibitem [{\citenamefont {Tchistiakov}(1996)}]{Tchistiakov1996}%
  \BibitemOpen
  \bibfield  {author} {\bibinfo {author} {\bibfnamefont {V.}~\bibnamefont
  {Tchistiakov}},\ }\href {\doibase
  https://doi.org/10.1016/0167-2789(95)00253-7} {\bibfield  {journal} {\bibinfo
   {journal} {Physica D: Nonlinear Phenomena}\ }\textbf {\bibinfo {volume}
  {91}},\ \bibinfo {pages} {67} (\bibinfo {year} {1996})}\BibitemShut {NoStop}%
\bibitem [{\citenamefont {Strogatz}(2015)}]{Strogatz2015}%
  \BibitemOpen
  \bibfield  {author} {\bibinfo {author} {\bibfnamefont {S.~H.}\ \bibnamefont
  {Strogatz}},\ }\href@noop {} {\emph {\bibinfo {title} {Nonlinear Dynamics and
  Chaos, 2nd Edition}}}\ (\bibinfo  {publisher} {Routledge Tazlor \& Francis
  Group},\ \bibinfo {address} {Oxford, UK},\ \bibinfo {year} {2015})\ p.\
  \bibinfo {pages} {532}\BibitemShut {NoStop}%
\bibitem [{\citenamefont {Feigenbaum}(1978)}]{Feigenbaum1978}%
  \BibitemOpen
  \bibfield  {author} {\bibinfo {author} {\bibfnamefont {M.~J.}\ \bibnamefont
  {Feigenbaum}},\ }\href {\doibase 10.1007/BF01020332} {\bibfield  {journal}
  {\bibinfo  {journal} {Journal of Statistical Physics}\ }\textbf {\bibinfo
  {volume} {19}},\ \bibinfo {pages} {25} (\bibinfo {year} {1978})}\BibitemShut
  {NoStop}%
\bibitem [{\citenamefont {Golubitsky}\ and\ \citenamefont
  {Stewart}(2015)}]{Golubitsky2015}%
  \BibitemOpen
  \bibfield  {author} {\bibinfo {author} {\bibfnamefont {M.}~\bibnamefont
  {Golubitsky}}\ and\ \bibinfo {author} {\bibfnamefont {I.}~\bibnamefont
  {Stewart}},\ }\href {\doibase 10.1063/1.4918595} {\bibfield  {journal}
  {\bibinfo  {journal} {Chaos: An Interdisciplinary Journal of Nonlinear
  Science}\ }\textbf {\bibinfo {volume} {25}},\ \bibinfo {pages} {097612}
  (\bibinfo {year} {2015})},\ \Eprint
  {http://arxiv.org/abs/https://doi.org/10.1063/1.4918595}
  {https://doi.org/10.1063/1.4918595} \BibitemShut {NoStop}%
\bibitem [{\citenamefont {van~der Weele}(1992)}]{vanderWeele1992}%
  \BibitemOpen
  \bibfield  {author} {\bibinfo {author} {\bibfnamefont {J.~P.}\ \bibnamefont
  {van~der Weele}},\ }\enquote {\bibinfo {title} {Symmetry breaking in the
  period doubling route to chaos},}\ in\ \href {\doibase
  10.1007/978-1-4615-3464-8_33} {\emph {\bibinfo {booktitle} {Chaotic Dynamics:
  Theory and Practice}}},\ \bibinfo {editor} {edited by\ \bibinfo {editor}
  {\bibfnamefont {T.}~\bibnamefont {Bountis}}}\ (\bibinfo  {publisher}
  {Springer US},\ \bibinfo {address} {Boston, MA},\ \bibinfo {year} {1992})\
  pp.\ \bibinfo {pages} {357--369}\BibitemShut {NoStop}%
\bibitem [{\citenamefont {Garc\'{i}a-Morales}\ and\ \citenamefont
  {Krischer}(2012)}]{GarciaMorales2012}%
  \BibitemOpen
  \bibfield  {author} {\bibinfo {author} {\bibfnamefont {V.}~\bibnamefont
  {Garc\'{i}a-Morales}}\ and\ \bibinfo {author} {\bibfnamefont
  {K.}~\bibnamefont {Krischer}},\ }\href {\doibase
  10.1080/00107514.2011.642554} {\bibfield  {journal} {\bibinfo  {journal}
  {Contemporary Physics}\ }\textbf {\bibinfo {volume} {53}},\ \bibinfo {pages}
  {79} (\bibinfo {year} {2012})},\ \Eprint
  {http://arxiv.org/abs/http://dx.doi.org/10.1080/00107514.2011.642554}
  {http://dx.doi.org/10.1080/00107514.2011.642554} \BibitemShut {NoStop}%
\bibitem [{\citenamefont {Battogtokh}\ and\ \citenamefont
  {Mikhailov}(1996)}]{Battogtokh1996}%
  \BibitemOpen
  \bibfield  {author} {\bibinfo {author} {\bibfnamefont {D.}~\bibnamefont
  {Battogtokh}}\ and\ \bibinfo {author} {\bibfnamefont {A.}~\bibnamefont
  {Mikhailov}},\ }\href {\doibase https://doi.org/10.1016/0167-2789(95)00232-4}
  {\bibfield  {journal} {\bibinfo  {journal} {Physica D: Nonlinear Phenomena}\
  }\textbf {\bibinfo {volume} {90}},\ \bibinfo {pages} {84} (\bibinfo {year}
  {1996})}\BibitemShut {NoStop}%
\bibitem [{\citenamefont {Cox}\ and\ \citenamefont {Matthews}(2002)}]{Cox2002}%
  \BibitemOpen
  \bibfield  {author} {\bibinfo {author} {\bibfnamefont {S.}~\bibnamefont
  {Cox}}\ and\ \bibinfo {author} {\bibfnamefont {P.}~\bibnamefont {Matthews}},\
  }\href {\doibase 10.1006/jcph.2002.6995} {\bibfield  {journal} {\bibinfo
  {journal} {Journal of Computational Physics}\ }\textbf {\bibinfo {volume}
  {176}},\ \bibinfo {pages} {430} (\bibinfo {year} {2002})}\BibitemShut
  {NoStop}%
\bibitem [{\citenamefont {Hunter}(2007)}]{Hunter2007}%
  \BibitemOpen
  \bibfield  {author} {\bibinfo {author} {\bibfnamefont {J.~D.}\ \bibnamefont
  {Hunter}},\ }\href {\doibase 10.1109/MCSE.2007.55} {\bibfield  {journal}
  {\bibinfo  {journal} {Computing In Science \& Engineering}\ }\textbf
  {\bibinfo {volume} {9}},\ \bibinfo {pages} {90} (\bibinfo {year}
  {2007})}\BibitemShut {NoStop}%
\end{thebibliography}%

\end{document}